\documentclass[twocolumn,aps,prb,showpacs,amsmath,amssymb, floatfix]{revtex4}
\usepackage[dvips]{graphicx}
\usepackage{dcolumn}
\usepackage{bm}
\usepackage{color}

\begin{document}
\title{Self Consistent Simulation of Quantum Transport and Magnetization Dynamics in Spin-Torque Based Devices}
\author{Sayeef Salahuddin}
\email{ssalahud@purdue.edu}
\author{Supriyo Datta}
\affiliation{School of Electrical and Computer Engineering  and NSF Network for Computational Nanotechnology (NCN), Purdue University, West Lafayette, IN-47907, USA.}
\date{\today}

\begin{abstract}
We present a self consistent solution of quantum transport, using the Non Equilibrium Green's Function method, and magnetization dynamics, using the Landau-Lifshitz-Gilbert formulation. We have applied this model to study `spin torque' induced magnetic switching in a device where the transport is ballistic and the free magnetic layer is sandwiched between two anti-parallel (AP) ferromagnetic contacts. We predict hysteretic current-voltage characteristics, at room temperature, with a sharp transition between the bistable states that can be used as a non-volatile memory. We further show that this AP penta layer device may allow significant reduction in the  switching current, thus facilitating integration of nanomagnets with electronic devices.
\end{abstract}

\maketitle
Successful integration of nanomagnets with electronic devices may enable the first generation of practical spintronic devices, which have so far been elusive due to stringent requirements such as low temperature and high magnetic field. It was predicted by Slonczewski \cite{slonczewski96:ref} and Berger \cite{berger96:ref} that magnetization of a nano magnet may be flipped by a spin polarized current through the so-called `spin torque' effect and this was later demonstrated experimentally \cite{kiselev03_nature:ref,katine00:ref}.  However, the early spin-torque systems were metal based that allowed only a small change in the magnetoresistance. In addition, metallic channels are difficult to integrate with CMOS technology. Recently a number of experiments have demonstrated current-induced magnetization switching in MgO based Tunneling Magneto Resistance (TMR) devices at (i) room temperature (ii) with a TMR ratio of more than 100\% and (iii) without any external magnetic field  \cite{kubota05:ref,parkin04_nature:ref}. Encouraged by these experimental results, here we explore theoretically a memory device based on current induced magnetization switching in the quantum transport regime. 

\begin{figure}[b]
	\centering
	\includegraphics[height=4cm]{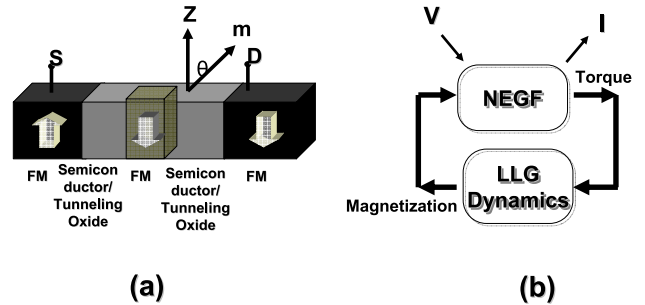}
  \caption{(a) A schematic showing the penta layer device. The free ferromagnetic layer is embedded inside the channel which is sandwiched between two `hard' ferromagnetic contacts.(b) A schematic showing the self-consistent nature of the transport problem. The magnetization dynamics and transport are mutually dependent on one another.}
	\label{FIG1}
\end{figure}

The device under consideration is shown in Fig. \ref{FIG1}. It consists of five layers. The two outer layers are `hard magnets' which act as spin polarized contacts. There is a soft magnetic layer inside the channel whose magnetization is affected by the current flow through the so-called `spin torque' effect. The channel can be a semiconductor \cite{harris05:ref} or a tunneling oxide \cite{parkin04_nature:ref}. Note that the contacts are arranged in an anti-parallel (AP) configuration. We have recently showed that in this configuration, the torque exerted by the injected electrons on a the nearby spin array (in this case the soft magnet) is maximum \cite{salahuddin06:ref}. A similar prediction was also made by Berger \cite{Berger_five_layer:ref} based on expansion/contraction of the Fermi surface. The possibility of an enhanced torque and therefore a lower switching current is our motivation for the penta-layer configuration instead of the conventional tri-layer geometry. 

In Fig. 1, the soft-magnet changes the transport through its interaction with the channel electrons, which in turn exert a torque on the magnet and try to rotate it from its equilibrium state. In this paper, we present a self-consistent solution of both these processes: the transport of channel electrons (through NEGF) and the magnetization dynamics of the free layer (through LLG equations)(see Fig. \ref{FIG1}(b)). Our calculations show clear hysteretic I-V suggesting possible use as a memory. Furthermore, we show that a penta-layer device with AP contact as shown in Fig.1(a) should exhibit a significant reduction in the switching current.

\begin{figure}[t]
	\centering
	\includegraphics[width=8cm]{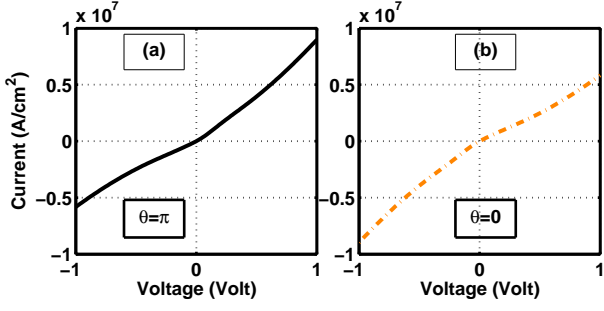}
  \caption{ Non self consistent (with magnetization dynamics) I-V characteristics of the the proposed device (a) With the soft magnet initially at $\theta=\pi$ position. THe  current is larger for positive bias. (b) With the soft magnet initially at $\theta=0$ position. The current is larger for negative bias.}
	\label{FIG3}
\end{figure}

\emph{Transport.-} Unlike the conventional metallic spin-torque systems, where the transport is predominantly diffusive, the transport in semiconductors or tunneling oxides is ballistic or quasi-ballistic. This necessitates a quantum description of the transport. We use the Non Equilibrium Green's Function (NEGF) method to treat the transport rigorously. The interaction between channel electrons and the ferromagnet is mediated through exchange and it is described by $H_I(\vec{r})=\sum_{R_j}J(\vec{r}-\vec{R_j})\vec{\sigma}.\vec{S_j}$, where, $r$ and $R_j$ are the spatial coordinates and $\sigma$ and $S_j$ are the spin operators for the channel electron and $j$-th spin in the soft-magnet. $J(\bar{r}-\bar{R_j})$ is the interaction constant between the channel electron and the j-th spin in the magnet. This interaction is taken into account through self energy ($\Sigma_s$), which is a function of the magnetization ($\vec{m}$), using the so-called self-consistent Born approximation \cite{datta-italian:ref}. In this method, the spin current flowing into the soft magnet is given by
\begin{equation}
\label{spin_current}
	\left[I_{spin}\right]=\int dE \frac{e}{\hbar}i\left[
																								Tr \left\{
																												G\Sigma^{in}_s-\Sigma^{in}_sG^\dagger-\Sigma_sG^n+G^n\Sigma^\dagger_s
																												\right\}
																						\right],
\end{equation}
where the trace is taken only over the space coordinates. Then $\left[I_{spin}\right]$ is a $2\times2$  matrix in the spin space. Here, $G$ denotes the Green's function. The torque exerted on the magnet is calculated from $\left[I_{spin}\right]$ by writing $T_i=Trace\left\{S_i\left[I_{spin}\right]\right\}$,where $i=\{x,y,z\}$. The total current, which is found from a similar expression as Eq. \ref{spin_current} with the self energy $\Sigma_s$ now replaced by the total self-energy $\Sigma$ \cite{datta_green:ref}, is shown in Fig.\ref{FIG3}(a),(b) for two different configuration of the magnetization. The nonlinear nature of the I-V can be intuitively understood by recognizing that the exchange interaction is minimum if the injected electrons and the magnetization has the same spin orientation \cite{salahuddin06:ref}. 

For the calculations, the Hamiltonian was written in the effective-mass approximation where the hopping parameter $t=\hbar^2/(2m^*a^2)$, $m^*=0.7m_e$ \cite{rippard_2002:ref} denoting the effective mass and $a$ being the lattice spacing. The interaction constant $J$ is assumed to be 0.01 eV \cite{mitchell_57:ref}. A sample set of parameters is $E_f=2 eV$ ; barrier-height=1.2 eV \cite{rippard_2002:ref}; barrier-width=1 nm and exchange splitting=1.2 eV \cite{himpsel_91:ref}. However, the barrier height, width and exchange splitting were artificially varied to get desired injection efficiency and TMR value. We also modified the perfect-contact self energy to read $\Sigma=-t^{'}exp(ika)$($t^{'}\neq t$ and $k$ is the momentum) to simulate the reflective nature of the contact (for details see \cite{datta_new:ref}). For plots 2-4, an injection efficiency of $70\%$ was assumed.

\emph{Magnetization Dynamics.-}The magnetization dynamics is simulated using the LLG equation:
\begin{equation}
\begin{split}
\left({1+\alpha^2}\right)\frac{\partial \vec{m}}{\partial t}=\gamma\left(\vec{m}\times\vec{H_{eff}}\right)
-\frac{\gamma\alpha}{m}\vec{m}\times\vec{m}\times\vec{H_{{eff}}}\\
+\text{Current Torque.}
\end{split}
\label{llg}
\end{equation}
Here, $\vec{m}$ is the magnetization of the soft magnet, $\gamma=17.6$ MHz/Oe is the gyro magnetic ratio and $\alpha$ is the Gilbert damping parameter. The $\vec{H_{eff}}=\vec{H_{ext}}+(2Ku_2/M_s)m_z\hat{z}-(2Ku_p/M_s)m_x\hat{x}$, where $H_{ext}$ is the externally applied magnetic field, $M_s$ is the saturation magnetization and $Ku_2$ and $Ku_p$ are the uniaxial and in-plane anisotropy constants respectively. The conventional LLG equation has to be solved with the current torque ($T_i$) that works as an additional source term. 
\begin{figure}[t]
	\centering
	\includegraphics[width=5.5cm]{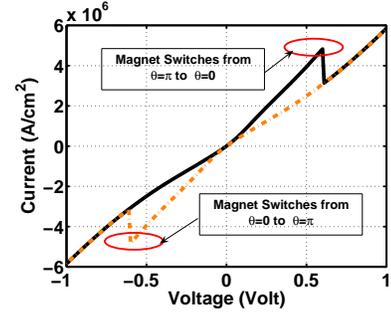}
  \caption{ The hysteretic I-V originating from a self-consistent solution of transport and LLG. At a certain bias, the current torque produced by the conduction electrons is strong enough to flip the magnet. These transition points are indicated in the figure. }
	\label{FIG6}
\end{figure}

\emph{Self-Consistency.-}Fig. \ref{FIG3} shows the situations when the transport and magnetization dynamics is independent of each other. This will change when Eqs. \ref{spin_current} and \ref{llg} are solved self-consistently. If we start from $\theta=\pi$ position, I-V curve follows the trend shown in Fig.\ref{FIG3}(a). However, once the torque exceeds the critical field (discussed later), the magnet switches abruptly. As a result I-V characteristics now follows that shown in Fig.\ref{FIG3}(b). This results in the hysteretic I-V shown in Fig. \ref{FIG6}. 
\begin{figure}[t]
	\centering
	\includegraphics[width=6cm]{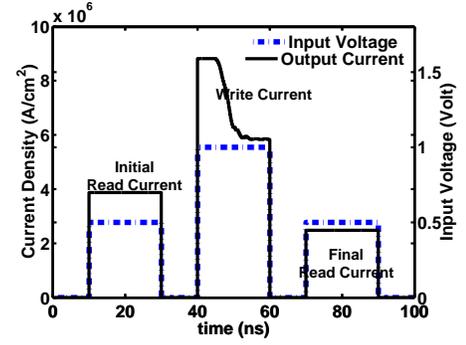}
  \caption{Response to a Read-Write-Read pulse. The write pulse switches the magnet from $\theta=\pi$ to $\theta=0$. The corresponding change in the current can be clearly seen during the write pulse.}
	\label{FIG7}
\end{figure}

Fig.\ref{FIG7} shows current flow in the device in response to Read-Write-Read pulse sequence. Here, we have used read pulse of 0.5 volt and write pulse of +1 volt. The soft magnet is initially in the $\theta=\pi$ position. The Write pulse switches it to $\theta=0$. Note the change in the current level in response to the Read pulse before and after applying the Write pulse.

\emph{Discussion.-}A question may be raised regarding the asymmetric I-V of Fig. \ref{FIG3}, which is not expected if one thinks about the device in Fig. \ref{FIG1}(a) as a series combination of two devices, one anti-parallel (AP) and one parallel (P). The device, however, is different from a mere series combination since the contact in the middle works as a mixing element for up and down spin electrons. The difference will be clear if one assumes 100\% injection efficiency. No current is expected to flow through the series combination of an AP and a P device. However, in our device, a current can still flow because the contact in the middle mixes the up and down spin channels. This `extra' current originating from `channel-mixing' gives the observed asymmetry in Fig. \ref{FIG3}.

Since electronic time constants are typically in the sub picosecond regime which is much faster than the magnetization dynamics (typically of the order of nano seconds), we have assumed that, for electronic transport, the magnetization dynamics is a quasi-static process \cite{self_cond_mat:ref}.

The switching is obtained by the torque component which is transverse to the magnetization of the soft-magnet. From Eq. \ref{llg}, considering average rate of change of energy, it can be shown that the magnitude of the torque required to induce switching is $\alpha\gamma\left(H_{ext}+H_k+H_p/2\right)$\cite{sun_prb:ref}, where $H_k=2Ku_2/M_s$ and $H_p=2Ku_p/M_s=4\pi M_s$. This then translates into a critical spin current magnitude of 
\begin{equation}
\label{spin_critical}
I_{spin}=\frac{2e}{\hbar}\alpha(M_sV)(H_{ext}+H_k+2\pi M_s).
\end{equation}
Here, $V$ is the volume of the free magnetic layer. Depending on the magnitudes of $\alpha$,$M_s$,$H_k$ and thickness, $d$, of the magnet, the spin current density to achieve switching varies from $10^5-10^6$ A/cm$^2$ (e.g for Co, using typical values $\alpha\sim0.01$,$H_k\sim 100$ Oe and $M_s=1.5\times10^3$ emu/cm$^3$ \cite{sun_prb:ref} and $d=2$ nm, the spin current density required  is roughly $10^6$ A/cm$^2$). Note that this requirement on spin current is completely determined by the magnetic properties of the free layer. The actual current density is typically another factor of 10-100 larger due to the additional coherent component of the current which does not require any spin-flip. Hence an important metric for critical current requirement is $r=I_{coherent}/I_{spin}$, which should be as small as possible. Intuitively, with AP contacts, the coherent current $I_{coherent}\propto t^22\alpha\beta$, where $t$ is the hopping matrix element, $\alpha$ is the majority(minority) density of states for the injecting contact and $\beta$ is the minority(majority) density of states of the drain contact. Similarly the spin-flip current $I_{sf}\propto J^2\left[\alpha^2(1-P_\alpha)-\beta^2P_\alpha\right]$, where $P_\alpha$ is the probability of a spin in the free layer to be in state $\alpha$ \cite{salahuddin06:ref}. It follows that 
\begin{equation}
\label{ap_ratio}
r_{AP}=\frac{I_{coherent}}{I_{sf}}|_{AP}=\frac{t^2}{J^2}\frac{1-P_c^2}{P_c+(\frac{1}{2}-P_\alpha)(1+P_c^2)},
\end{equation}
where, $P_c=(\alpha-\beta)/(\alpha+\beta)$ indicates the degree of contact polarization. These approximate analytical expressions (Eqs. \ref{ap_ratio}) agree quite well with detailed NEGF calculations described above. The $I_{coherent}$ and $I_{spin}$ can be found respectively from the symmetric and asymmetric portion of the non-linear I-V shown in Fig. 2. Fig. 5 shows the variation of $g=r_{tri-layer}/t_{AP-penta-layer}$ with $P_c$. The plot shows that $g>>1$ for reasonable values of $Pc$, indicating a lower switching current for the penta-layer device.  
\begin{figure}[t]
	\centering
	\includegraphics[width=6cm]{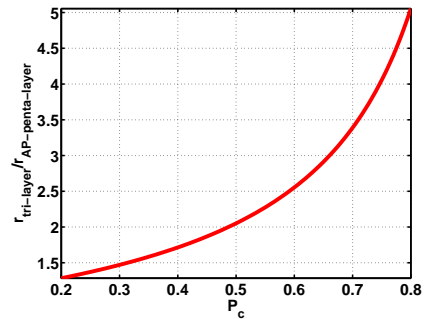}
  \caption{Variation of the ratio of $I_{coherent}/I_{spin}$ for a 3 layer and AP penta layer spin-torque device showing the possible reduction of switching current for the AP penta layer device compared to the 3-layer device.}
	\label{FIG8}
\end{figure}
Recent experiments on AP penta-layer devices \cite{fuchs:ref, huai:ref, meng:ref} have shown similar reduction of switching current compared to 3-layer devices. These experiments seem to follow the general trends of Fig. 5 as the reduction factor is seen to increase with increasing TMR (see Fig. 4 of \cite{huai:ref}). A detailed study of the dependence of the reduction factor on material parameters is beyond the scope of this paper.   



The sharp transition between HIGH and LOW states in Fig. 3 arises from the bistable nature of the solutions to the LLG equation in the absence of any external field perpendicular to the easy axis. The intrinsic speed depends on $\omega=\gamma B$ where $B$ can be roughly estimated as $B\sim \hbar T/(2\mu_B)$. A higher speed will require higher current density.

In conclusion, we have shown a scheme for calculating the `spin-current' and the corresponding torque directly from transport parameters within the framework of NEGF formalism. A non-linear I-V is predicted for AP penta-layer devices. Experimental observation of this non-linearity (which can also be detected as steps or peaks in respectively the first and second derivative of the I-V \cite{salahuddin06:ref}) would provide strong confirmation of our approach. We have further coupled the transport formalism with the phenomenological magnetization dynamics (LLG equation). Our self consistent simulation of NEGF-LLG equations show clear hysteretic switching behavior, which is a direct consequence of the non-linearity described above. Finally, we have shown that the switching current for AP penta layer devices can be significantly lower than the conventional 3 layer devices.


This work was supported by the MARCO focus center for Materials, Structure and Devices. 
%
\bibliography{memory}

\begin{thebibliography}{20}
\expandafter\ifx\csname natexlab\endcsname\relax\def\natexlab#1{#1}\fi
\expandafter\ifx\csname bibnamefont\endcsname\relax
  \def\bibnamefont#1{#1}\fi
\expandafter\ifx\csname bibfnamefont\endcsname\relax
  \def\bibfnamefont#1{#1}\fi
\expandafter\ifx\csname citenamefont\endcsname\relax
  \def\citenamefont#1{#1}\fi
\expandafter\ifx\csname url\endcsname\relax
  \def\url#1{\texttt{#1}}\fi
\expandafter\ifx\csname urlprefix\endcsname\relax\def\urlprefix{URL }\fi
\providecommand{\bibinfo}[2]{#2}
\providecommand{\eprint}[2][]{\url{#2}}

\bibitem[{\citenamefont{Slonczewski}(1996)}]{slonczewski96:ref}
\bibinfo{author}{\bibfnamefont{J.~C.} \bibnamefont{Slonczewski}},
  \bibinfo{journal}{Journal of Magnetism and Magnetic Materials}
  \textbf{\bibinfo{volume}{159}}, \bibinfo{pages}{L1} (\bibinfo{year}{1996}).

\bibitem[{\citenamefont{Berger}(1996)}]{berger96:ref}
\bibinfo{author}{\bibfnamefont{L.}~\bibnamefont{Berger}},
  \bibinfo{journal}{Physical Review B} \textbf{\bibinfo{volume}{54}},
  \bibinfo{pages}{9353} (\bibinfo{year}{1996}).

\bibitem[{\citenamefont{Kiselev et~al.}(2003)\citenamefont{Kiselev, Sankey,
  Krivorotov, Emley, Schoelkopf, Buhrman, and Ralph}}]{kiselev03_nature:ref}
\bibinfo{author}{\bibfnamefont{S.~I.} \bibnamefont{Kiselev}},
  \bibinfo{author}{\bibfnamefont{J.~C.} \bibnamefont{Sankey}},
  \bibinfo{author}{\bibfnamefont{I.~N.} \bibnamefont{Krivorotov}},
  \bibinfo{author}{\bibfnamefont{N.~C.} \bibnamefont{Emley}},
  \bibinfo{author}{\bibfnamefont{R.~J.} \bibnamefont{Schoelkopf}},
  \bibinfo{author}{\bibfnamefont{R.~A.} \bibnamefont{Buhrman}},
  \bibnamefont{and} \bibinfo{author}{\bibfnamefont{D.~C.} \bibnamefont{Ralph}},
  \bibinfo{journal}{Nature} \textbf{\bibinfo{volume}{425}},
  \bibinfo{pages}{380} (\bibinfo{year}{2003}).

\bibitem[{\citenamefont{Katine et~al.}(2000)\citenamefont{Katine, Albert,
  Buhrman, Myers, and Ralph}}]{katine00:ref}
\bibinfo{author}{\bibfnamefont{J.~A.} \bibnamefont{Katine}},
  \bibinfo{author}{\bibfnamefont{F.~J.} \bibnamefont{Albert}},
  \bibinfo{author}{\bibfnamefont{R.~A.} \bibnamefont{Buhrman}},
  \bibinfo{author}{\bibfnamefont{E.~B.} \bibnamefont{Myers}}, \bibnamefont{and}
  \bibinfo{author}{\bibfnamefont{D.~C.} \bibnamefont{Ralph}},
  \bibinfo{journal}{Physical Review Letters} \textbf{\bibinfo{volume}{84}},
  \bibinfo{pages}{3149} (\bibinfo{year}{2000}).

\bibitem[{\citenamefont{Kubota et~al.}(2005)\citenamefont{Kubota, Fukushima,
  Ootani, Yuasa, Ando, Maehara, Tsunekawa, Djayaprawira, Watanabe, and
  Suzuki}}]{kubota05:ref}
\bibinfo{author}{\bibfnamefont{H.}~\bibnamefont{Kubota}},
  \bibinfo{author}{\bibfnamefont{A.}~\bibnamefont{Fukushima}},
  \bibinfo{author}{\bibfnamefont{Y.}~\bibnamefont{Ootani}},
  \bibinfo{author}{\bibfnamefont{S.}~\bibnamefont{Yuasa}},
  \bibinfo{author}{\bibfnamefont{K.}~\bibnamefont{Ando}},
  \bibinfo{author}{\bibfnamefont{H.}~\bibnamefont{Maehara}},
  \bibinfo{author}{\bibfnamefont{K.}~\bibnamefont{Tsunekawa}},
  \bibinfo{author}{\bibfnamefont{D.~D.} \bibnamefont{Djayaprawira}},
  \bibinfo{author}{\bibfnamefont{N.}~\bibnamefont{Watanabe}}, \bibnamefont{and}
  \bibinfo{author}{\bibfnamefont{Y.}~\bibnamefont{Suzuki}},
  \bibinfo{journal}{Japanese Journal of Applied Physics}
  \textbf{\bibinfo{volume}{44}}, \bibinfo{pages}{L1237} (\bibinfo{year}{2005}).

\bibitem[{\citenamefont{Parkin et~al.}(2004)\citenamefont{Parkin, Kaiser,
  Panchula, Rice, Hughes, Samant, and Yang}}]{parkin04_nature:ref}
\bibinfo{author}{\bibfnamefont{S.~S.~P.} \bibnamefont{Parkin}},
  \bibinfo{author}{\bibfnamefont{C.}~\bibnamefont{Kaiser}},
  \bibinfo{author}{\bibfnamefont{A.}~\bibnamefont{Panchula}},
  \bibinfo{author}{\bibfnamefont{P.~M.} \bibnamefont{Rice}},
  \bibinfo{author}{\bibfnamefont{B.}~\bibnamefont{Hughes}},
  \bibinfo{author}{\bibfnamefont{M.}~\bibnamefont{Samant}}, \bibnamefont{and}
  \bibinfo{author}{\bibfnamefont{S.~H.} \bibnamefont{Yang}},
  \bibinfo{journal}{Nature Materials} \textbf{\bibinfo{volume}{3}},
  \bibinfo{pages}{862} (\bibinfo{year}{2004}).

\bibitem[{\citenamefont{Jiang et~al.}(2005)\citenamefont{Jiang, Wang, Shelby,
  Macfarlane, Bank, Harris, and Parkin}}]{harris05:ref}
\bibinfo{author}{\bibfnamefont{X.}~\bibnamefont{Jiang}},
  \bibinfo{author}{\bibfnamefont{R.}~\bibnamefont{Wang}},
  \bibinfo{author}{\bibfnamefont{R.~M.} \bibnamefont{Shelby}},
  \bibinfo{author}{\bibfnamefont{R.~M.} \bibnamefont{Macfarlane}},
  \bibinfo{author}{\bibfnamefont{S.~R.} \bibnamefont{Bank}},
  \bibinfo{author}{\bibfnamefont{J.~S.} \bibnamefont{Harris}},
  \bibnamefont{and} \bibinfo{author}{\bibfnamefont{S.~S.~P.}
  \bibnamefont{Parkin}}, \bibinfo{journal}{Physical Review Letters}
  \textbf{\bibinfo{volume}{94}}, \bibinfo{pages}{056601}
  (\bibinfo{year}{2005}).

\bibitem[{\citenamefont{Salahuddin and
  Datta}(2006{\natexlab{a}})}]{salahuddin06:ref}
\bibinfo{author}{\bibfnamefont{S.}~\bibnamefont{Salahuddin}} \bibnamefont{and}
  \bibinfo{author}{\bibfnamefont{S.}~\bibnamefont{Datta}},
  \bibinfo{journal}{Physical Review B} \textbf{\bibinfo{volume}{73}},
  \bibinfo{pages}{081301R} (\bibinfo{year}{2006}{\natexlab{a}}).

\bibitem[{\citenamefont{Berger}(2003)}]{Berger_five_layer:ref}
\bibinfo{author}{\bibfnamefont{L.}~\bibnamefont{Berger}},
  \bibinfo{journal}{Journal of Applied Physics} \textbf{\bibinfo{volume}{93}},
  \bibinfo{pages}{7693} (\bibinfo{year}{2003}).

\bibitem[{\citenamefont{Datta}(2005{\natexlab{a}})}]{datta-italian:ref}
\bibinfo{author}{\bibfnamefont{S.}~\bibnamefont{Datta}},
  \bibinfo{journal}{Proceedings of the Intenational School of Physics Enrico
  Fermi, Italiana di Fisica}  (\bibinfo{year}{2005}{\natexlab{a}}).

\bibitem[{\citenamefont{Datta}(1995)}]{datta_green:ref}
\bibinfo{author}{\bibfnamefont{S.}~\bibnamefont{Datta}},
  \emph{\bibinfo{title}{Electronic Transport in Mesoscopic Systems}}
  (\bibinfo{publisher}{Cambridge University Press}, \bibinfo{year}{1995}).

\bibitem[{\citenamefont{Rippard et~al.}(2002)\citenamefont{Rippard, Perrella,
  Albert, and Buhrman}}]{rippard_2002:ref}
\bibinfo{author}{\bibfnamefont{W.~H.} \bibnamefont{Rippard}},
  \bibinfo{author}{\bibfnamefont{A.~C.} \bibnamefont{Perrella}},
  \bibinfo{author}{\bibfnamefont{F.~J.} \bibnamefont{Albert}},
  \bibnamefont{and} \bibinfo{author}{\bibfnamefont{R.~A.}
  \bibnamefont{Buhrman}}, \bibinfo{journal}{Physical Review Letters}
  \textbf{\bibinfo{volume}{88}}, \bibinfo{pages}{046805}
  (\bibinfo{year}{2002}).

\bibitem[{\citenamefont{Mitchell}(1957)}]{mitchell_57:ref}
\bibinfo{author}{\bibfnamefont{A.~H.} \bibnamefont{Mitchell}},
  \bibinfo{journal}{Physical Review} \textbf{\bibinfo{volume}{105}},
  \bibinfo{pages}{1439} (\bibinfo{year}{1957}).

\bibitem[{\citenamefont{Himpsel}(1991)}]{himpsel_91:ref}
\bibinfo{author}{\bibfnamefont{F.~J.} \bibnamefont{Himpsel}},
  \bibinfo{journal}{Physical Review Letters} \textbf{\bibinfo{volume}{67}},
  \bibinfo{pages}{2363} (\bibinfo{year}{1991}).

\bibitem[{\citenamefont{Datta}(2005{\natexlab{b}})}]{datta_new:ref}
\bibinfo{author}{\bibfnamefont{S.}~\bibnamefont{Datta}},
  \emph{\bibinfo{title}{Quantum Transport: Atom to Transistor}}
  (\bibinfo{publisher}{Cambridge University Press},
  \bibinfo{year}{2005}{\natexlab{b}}).

\bibitem[{\citenamefont{Salahuddin and
  Datta}(2006{\natexlab{b}})}]{self_cond_mat:ref}
\bibinfo{author}{\bibfnamefont{S.}~\bibnamefont{Salahuddin}} \bibnamefont{and}
  \bibinfo{author}{\bibfnamefont{S.}~\bibnamefont{Datta}},
  \bibinfo{journal}{cond-mat/0606648}  (\bibinfo{year}{2006}{\natexlab{b}}).

\bibitem[{\citenamefont{Sun}(2000)}]{sun_prb:ref}
\bibinfo{author}{\bibfnamefont{J.~Z.} \bibnamefont{Sun}},
  \bibinfo{journal}{Physical Review B} \textbf{\bibinfo{volume}{62}},
  \bibinfo{pages}{570} (\bibinfo{year}{2000}).

\bibitem[{\citenamefont{Fuchs et~al.}(2005)\citenamefont{Fuchs, Krivorotov,
  Braganca, Emley, Garcia, Ralph, and Buhrman}}]{fuchs:ref}
\bibinfo{author}{\bibfnamefont{G.~D.} \bibnamefont{Fuchs}},
  \bibinfo{author}{\bibfnamefont{I.~N.} \bibnamefont{Krivorotov}},
  \bibinfo{author}{\bibfnamefont{P.~M.} \bibnamefont{Braganca}},
  \bibinfo{author}{\bibfnamefont{N.~C.} \bibnamefont{Emley}},
  \bibinfo{author}{\bibfnamefont{A.~G.~F.} \bibnamefont{Garcia}},
  \bibinfo{author}{\bibfnamefont{D.~C.} \bibnamefont{Ralph}}, \bibnamefont{and}
  \bibinfo{author}{\bibfnamefont{R.~A.} \bibnamefont{Buhrman}},
  \bibinfo{journal}{Applied Physics Letters} \textbf{\bibinfo{volume}{86}},
  \bibinfo{pages}{152509} (\bibinfo{year}{2005}).

\bibitem[{\citenamefont{Huai et~al.}(2005)\citenamefont{Huai, Pakala, Diao, and
  Ding}}]{huai:ref}
\bibinfo{author}{\bibfnamefont{Y.~M.} \bibnamefont{Huai}},
  \bibinfo{author}{\bibfnamefont{M.}~\bibnamefont{Pakala}},
  \bibinfo{author}{\bibfnamefont{Z.~T.} \bibnamefont{Diao}}, \bibnamefont{and}
  \bibinfo{author}{\bibfnamefont{Y.~F.} \bibnamefont{Ding}},
  \bibinfo{journal}{Applied Physics Letters} \textbf{\bibinfo{volume}{87}},
  \bibinfo{pages}{222510} (\bibinfo{year}{2005}).

\bibitem[{\citenamefont{Meng et~al.}(2006)\citenamefont{Meng, Wang, and
  J.P.}}]{meng:ref}
\bibinfo{author}{\bibfnamefont{H.}~\bibnamefont{Meng}},
  \bibinfo{author}{\bibfnamefont{J.}~\bibnamefont{Wang}}, \bibnamefont{and}
  \bibinfo{author}{\bibfnamefont{W.}~\bibnamefont{J.P.}},
  \bibinfo{journal}{Applied Physics Letters} \textbf{\bibinfo{volume}{88}},
  \bibinfo{pages}{082504} (\bibinfo{year}{2006}).

\end{thebibliography}
\end{document}